\documentclass[prd,aps,preprint,amsmath,amssymb,eshowkeys,nofootinbib]{revtex4-1}
\usepackage[mathscr]{euscript}
\usepackage{dcolumn}
\usepackage{multirow}
\usepackage{bm}
\usepackage{graphicx}
\usepackage{subfigure}
\usepackage{rotating}
\usepackage{amsmath,amssymb,amsthm}
\usepackage[colorlinks=true,linkcolor=red,urlcolor=green,citecolor=green]{hyperref}
\newcommand{\bea}{\begin{eqnarray}}
\newcommand{\eea}{\end{eqnarray}}
\newcommand{\beq}{\begin{equation}}
\newcommand{\eeq}{\end{equation}}

\def\/{\over}

\begin{document}

\title{Agegraphic dark energy from entropy of the anti-de Sitter black hole}
\author{Qihong Huang$^{1}$\footnote{Corresponding author: huangqihongzynu@163.com}, Yang Liu$^{2}$ and He Huang$^{3}$}

\affiliation{
$^1$ School of Physics and Electronic Science, Zunyi Normal University, Zunyi, Guizhou 563006, China\\
$^2$ Purple Mountain Observatory, Chinese Academy of Sciences, Nanjing 210023, China\\
$^3$ College of Mechanical and Electrical Engineering, Jiaxing Nanhu University, Jiaxing, Zhejiang 314001, China
}

\begin{abstract}
In this paper, we analyze the agegraphic dark energy from the entropy of the anti-de Sitter black hole using the age of the universe as the IR cutoff. We constrain its parameter with the Pantheon+ Type Ia supernova sample and observational Hubble parameter data, finding that the Akaike Information Criterion cannot effectively distinguish this model from the standard $\Lambda$CDM model. The present value of Hubble constant $H_{0}$ and the model parameter $b^{2}$ are constrained to $H_{0}=67.7 \pm 1.8$ and $b^{2}=0.303^{+0.019}_{-0.024}$. This model realizes the whole evolution of the universe, including the late-time accelerated expansion. Although it asymptotically approaches the standard $\Lambda$CDM model in the future, statefinder analysis shows that late-time deviations allow the two models to be distinguished.
\end{abstract}

\maketitle

\section{Introduction}

Observations confirm that the universe is currently undergoing accelerated expansion~\cite{Perlmutter1999, Riess1998, Spergel2003, Spergel2007, Tegmark2004, Eisenstein2005}, yet the underlying cause remains one of cosmology's central mysteries. To account for this phenomenon, dark energy has been proposed, with the simplest candidate being the $\Lambda$CDM model. Although $\Lambda$CDM agrees well with present observations~\cite{Planck2020}, it suffers from fine-tuning~\cite{Weinberg1989} and coincidence problems~\cite{Steinhardt1999}. This has motivated a variety of alternative dark energy models, including quintessence~\cite{Wetterich1988,Ratra1988,Caldwell1998}, quintom~\cite{Feng2005,Feng2006,Guo2005}, phantom~\cite{Caldwell2002,Caldwell2003}, k-essence~\cite{Chiba2000, Armendariz2001}, agegraphic dark energy (ADE)~\cite{Cai2007,Wei2007,Wei2008}, and holographic dark energy (HDE)~\cite{Cohen1999,Hsu2004,Li2004,Wang2017}.

A viable dark energy model must account not only for the universe's present accelerated expansion but also for its full evolutionary history, including the radiation-dominated era, the matter-dominated era, and the final dark energy-dominated epoch. To probe such evolutionary dynamics, phase-space analysis is employed, where critical points represent different cosmic stages: stable attractors correspond to dark energy dominance, while other fixed points trace earlier epochs~\cite{Bahamonde2018}. This approach has been widely applied to HDE models with considerable success~\cite{Setare2009, Liu2010, Banerjee2015, Mahata2015, Mishra2019, Bargach2019, Tita2024, Huang2019a, Ebrahimi2020, Astashenok2023, Huang2021, Srivastava2021}. Although dynamical system analysis provides a powerful method for examining the theoretical viability of dark energy models, a conclusive assessment requires comparison with precise observational data, such as the Pantheon+ Type Ia supernova data (SN Ia) sample~\cite{Riess2022, Brout2022, Scolnic2022} and observational Hubble parameter data (OHD)~\cite{Cao2022}. SN Ia, which acts as standardizable candles, allows for precise distance measurements and was instrumental in revealing the late-time cosmic acceleration. Similarly, OHD, obtained from the differential ages of galaxies or baryon acoustic oscillations, provides a direct probe of the expansion history. Comparison of dark energy models with these high-precision datasets enables us to not only test the consistency of the theoretical framework but also to place tight constraints on the key parameters. Consequently, combing Type Ia supernova data and OHD datesets has become a standard and powerful approach for constraining the parameters of cosmological models and is widely applied in cosmological research~\cite{Wang2025, Liu2024, Liu2024b, Arora2024, Oliveros2024, Wang2023, Mukherjee2022a, Mukherjee2022, Cao2021a, Pacif2021, Cao2021, Akarsu2020, Yang2020, Liu2019, Jimenez2016, Gong2013, Su2011, Gong2010, Wu2010a, Gong2006, Wu2006}.

The holographic dark energy (HDE) model~\cite{Hsu2004, Horvat2004, Li2004}, motivated by the holographic principle that bounds a region's entropy by its surface area~\cite{Witten1998, Bousso2002}, is a leading candidate for dark energy. The formulation of HDE models is based on the horizon entropy, and an alteration to this entropy gives rise to different models. This relationship exists because the specific form of the entropy directly determines the HDE density. For instance, the choice of the Bekenstein--Hawking entropy as the horizon entropy together with the Hubble horizon as the infrared (IR) cutoff yields the original and standard HDE model~\cite{Cohen1999,Hsu2004,Li2004}. Although the original HDE model was a significant proposal, it ultimately fails to account for the entire cosmic evolution~\cite{Li2004, Wang2017}. To address this limitation, various modifications have been proposed, including alternative IR cutoffs and interactions between dark matter and dark energy, giving rise to a range of extended HDE models~\cite{Wang2017}. Among these extensions, the ADE model stands out by utilizing the age of the universe as the IR cutoff~\cite{Cai2007}, offering a distinct approach to characterizing the dark energy evolution. The model is constructed from the K\'{a}rolyh\'{a}zy relation and the time--energy uncertainty principle, which gives it a similar energy density structure to that of HDE. While the original ADE avoids the causality problem and yields late-time acceleration, it fails to reproduce the matter-dominated era and mimic the cosmological constant at late stages. These shortcomings were later addressed by replacing cosmic time with conformal time as the IR cutoff, leading to the new ADE model~\cite{Wei2008}. As a result, ADE attracts extensive theoretical studies~\cite{HuangH2021, Pankaj2022, Mangoudehi2022, Sharma2020, HuangH2022, Sheykhi2023, Sharma2021, Ravanpak2019, Zadeh2019, Kumar2017, Saha2016, Zhang2014, Zhang2013, Farajollahi2012a, Li2013, Saaidi2012, Farajollahi2012, Sun2011a, Li2011, Lemets2011, Sun2011, Wei2008a, Wei2009}. 

The AdS/CFT correspondence~\cite{Maldacena1998} is a foundational conjecture in theoretical physics, which posits a remarkable duality between a gravitational theory in a higher dimensional anti-de Sitter (AdS) space and a conformal field theory (CFT) residing on its lower dimensional boundary. This duality provides a powerful framework for studying strongly coupled quantum field theories by translating them into more manageable problems in classical gravity. It has offered profound insights into a wide range of physical systems, including the properties of quark--gluon plasma~\cite{Policastro2001}, the mechanism of high temperature superconductivity~\cite{Hartnoll2008}, and the modified Friedmann equations in cosmology~\cite{Apostolopoulos2009, Khimphun2021}. Following in these footsteps, researchers have recently constructed a new HDE model using entropy calculations from AdS black holes, incorporating the Hubble horizon as the IR cutoff~\cite{Nakarachinda2022}. This model not only achieves late-time acceleration but also consistently describes the universe's full evolutionary history, including natural realizations of slow-roll inflation and reheating~\cite{Huang2025a}. Inspired by these results, we construct an ADE model using the entropy of AdS black holes with the universe's age as the IR cutoff (ADEADS) and investigate four key inquiries: (i) whether ADEADS's parameter space is consistent with the SN Ia and OHD datasets, (ii) whether it can realize late-time acceleration, (iii) whether it can reproduce the universe's complete evolution, and (iv) whether it can be differentiated from the standard $\Lambda$CDM model.

This paper is structured as follows: Section~\ref{sec:2} introduces the ADEADS model, and Section~\ref{sec:3} constrains its parameters. Section~\ref{sec:4} analyzes the universe's evolution within this framework, while Section~\ref{sec:5} examines the corresponding Hubble diagram. Section~\ref{sec:6} discusses the model's dynamical behavior, and Section~\ref{sec:7} employs statefinder diagnostics to distinguish ADEADS from $\Lambda$CDM. Finally, Section~\ref{sec:8} summarizes our main conclusions.

\section{Model} \label{sec:2}

Using the entropy of the AdS black hole, the HDE density takes the form~\cite{Nakarachinda2022}
\begin{equation}
\rho_{L}=\frac{3}{\kappa^{2}} b^{2} \Big(\frac{1}{L^{2}}+\Lambda\Big),
\end{equation}
where $\kappa^{2}=8\pi G$, $b$ is a dimensionless parameter and takes the value in the range $0<b^{2}<1$, and $\Lambda$ denotes the cosmological constant, which critically enables accelerated cosmic expansion during the universe's later evolutionary phases. The cosmological constant $\Lambda$ in the ADEADS framework originates intrinsically from the thermodynamic geometry of the anti-de Sitter black hole, contrasting with its phenomenological introduction in the $\Lambda$CDM model. When applying the first law of thermodynamics and Cohen's holographic bound to the mass of the AdS black hole, the energy density is derived. The term $\frac{3}{\kappa^{2}}b^{2}\Lambda$ arises directly from the AdS geometry, establishing a fundamental link between black hole thermodynamics and the cosmic acceleration, where a positive $\Lambda$ asymptotically drives de Sitter expansion despite the underlying gravitational context being anti-de Sitter. This reconceptualization, which embeds $\Lambda$ within a holographic principle, offers a quantum-gravitational perspective through the AdS/CFT duality. When $\Lambda$ vanishes, the energy density of the ADEADS framework reduces to that of the original ADE model~\cite{Cai2007}, thus demonstrating the latter's inherent limitations in replicating dark energy dynamics during the late-time cosmic evolution. When choosing the age of the universe
\begin{equation}
T=\int_{0}^{a} dt=\int_{0}^{a}\frac{da}{a H},
\end{equation}
as the IR cutoff, where $a$ is the scale factor, $t$ denotes cosmic time, and $H$ represents the Hubble parameter, and the energy density of ADEADS takes the form
\begin{equation}
\rho_{de}=\frac{3}{\kappa^{2}} b^{2} \Big(\frac{1}{T^{2}}+\Lambda\Big),\label{T0}
\end{equation}

Considering a homogeneous and isotropic Friedmann--Robertson--Walker universe with the line element
\begin{equation}
ds^{2}=-dt^{2}+a^{2}(t)(dr^{2}+r^{2}d\Omega^{2}),
\end{equation}
the Friedmann equation is
\begin{equation}
H^{2}=\frac{\kappa^{2}}{3}\big(\rho_{r}+\rho_{m}+\rho_{de}\big),\label{H2}
\end{equation}
where $\rho_{r}$, $\rho_{m}$, and $\rho_{de}$ represent the energy densities of pressureless matter, radiation, and ADEADS, respectively, and their conservation equations are
\begin{eqnarray}
&& \dot{\rho}_{r}+3H\rho_{r}=0,\label{rhor}\\
&& \dot{\rho}_{m}+3H\rho_{m}=0,\label{rhom}\\
&& \dot{\rho}_{de}+3H(1+\omega_{de})\rho_{de}=0,\label{rhode}
\end{eqnarray}
with $\omega_{de}$ as the equation of state parameter and satisfying $p_{de}=\omega_{de}\rho_{de}$. To analyze the dynamical evolution of the universe, we introduce three dimensionless variables
\begin{equation}
\Omega_{r}=\frac{\kappa^{2}\rho_{r}}{3H^{2}}, \quad \Omega_{m}=\frac{\kappa^{2}\rho_{m}}{3H^{2}}, \quad \Omega_{de}=\frac{\kappa^{2}\rho_{de}}{3H^{2}},
\end{equation}
and rewrite the Friedmann equation~(\ref{H2}) as
\begin{equation}
\Omega_{r}+\Omega_{m}+\Omega_{de}=1.\label{O1}
\end{equation}
Combining Equations~(\ref{H2})--(\ref{rhode}) and~(\ref{O1}), we obtain
\begin{equation}
\frac{\dot{H}}{H^{2}}=\frac{1}{2}\Big[\Omega_{m}+(1-3\omega_{de})\Omega_{de}\Big]-2,\label{HH2}
\end{equation}
and the deceleration parameter $q$ is defined as
\begin{equation}
q=-1-\frac{\dot{H}}{H^{2}}.\label{q0}
\end{equation}

Using Equations~(\ref{rhom}),~(\ref{rhode}),~(\ref{O1}), and~(\ref{HH2}) and the definition $'=\frac{d}{d(\ln a)}$, the automatic dynamical equations for this system can be written as
\begin{eqnarray}
&& \Omega'_{m}=[(3\omega_{de}-1)\Omega_{de}-\Omega_{m}+1]\Omega_{m},\label{Omm}\\
&& \Omega'_{de}=[(3\omega_{de}-1)(\Omega_{de}-1)-\Omega_{m}]\Omega_{de},\label{Omde}\\
&& \xi'=\Big[ 2-\frac{1}{2}(\Omega_{m}+(1-3\omega_{de})\Omega_{de}) \Big]\xi-\xi^{2},\label{OHT}
\end{eqnarray}
with
\begin{equation}
\omega_{de}=-1+\frac{2}{3}b^{2}\frac{\xi^{3}}{\Omega_{de}},\label{w0}
\end{equation}
and
\begin{equation}
\xi=\frac{1}{H T}.
\end{equation}

\section{Observational Constraints} \label{sec:3}

In this section, we utilize SN Ia sample covering $z \in [0.001, 2.261]$~\cite{Riess2022, Brout2022, Scolnic2022} and OHD spanning the redshift range $z \in [0.07, 1.965]$ collected by~\cite{Cao2022} to constrain the model parameters. We exclude SN Ia with redshifts below $0.01$ since their redshift measurements are strongly affected by unmodeled peculiar velocities~\cite{Brout2022}. In the ADEADS model, the expansion rate function can be written as
\begin{equation}
E(z)=\frac{H(z)}{H_{0}}=\sqrt{\Omega_{r,0}(1+z)^{4}+\Omega_{m,0}(1+z)^{3}+\Omega_{de,0}(1+z)^{3(\omega_{de}+1)}},
\end{equation}
where $\Omega_{r,0}+\Omega_{m,0}+\Omega_{de,0}=1$. The theoretical value of the observed quantity of SN Ia, the apparent magnitude $m_{th}$, is expressed as
\begin{equation}
m_{th}=5 \log_{10} \Big( \frac{D_{L}(z)}{Mpc} \Big) + 25 + M,
\end{equation}
where $M$ denotes the absolute magnitude of SN Ia and $D_{L}(z)$ is the luminosity distance, which takes the form
\begin{equation}
D_{L}(z)=c(1+z)\int^{z}_{0}\frac{dz}{H(z)}.
\end{equation}

To estimate the model parameters, we perform Markov Chain Monte Carlo (MCMC) sampling using the emcee library in Python 3.13.0. Since there is strong degeneracy between $b^{2}$ and $\alpha=\frac{\Lambda}{H^{2}_{0}}$, they cannot be effectively constrained at the same time. Thus, we calculate $\Lambda$ using the current observational data from the Planck 2018 results~\cite{Planck2020} and constrain the remaining four parameters $\{ H_{0}, \Omega_{m,0}, \Omega_{de,0}, b^{2} \}$. The total log-likelihood is defined as
\begin{equation}
\ln(\mathcal{L}_{total})=-\frac{1}{2}\chi^{2}_{total}+const.,
\end{equation}
with
\begin{equation}
\chi^{2}_{total}=\chi^{2}_{SN}+\chi^{2}_{OHD}.
\end{equation}
Here, the $\chi^{2}$ term for the SN Ia sample is calculated by
\begin{equation}
\chi^{2}_{SN}=\left( \hat{m}_{obs} - m_{th} \right)^\dag C_{SN}^{-1} \left( \hat{m}_{obs} - m_{th} \right),
\end{equation}
where $\hat{m}_{obs}$ represents the array of observed corrected apparent magnitude, and $C_{SN}$ is the corresponding covariance matrix. For the OHD sample, the $\chi^{2}$ term is given by 
\begin{equation}
\chi^{2}_{OHD} = \sum_{i=1}^{N_{OHD}} \left(\frac{H_{obs,i}-H_{th}(z_i)}{\sigma_{OHD,i}}\right)^2
\end{equation}
where $H_{obs,i}$ and $\sigma_{OHD,i}$ are the $i$-th observed value and its standard deviation, respectively, and $N_{OHD}$ is the total number of OHD data points.

\begin{figure*}[htp]
\begin{center}
\includegraphics[width=0.9\textwidth]{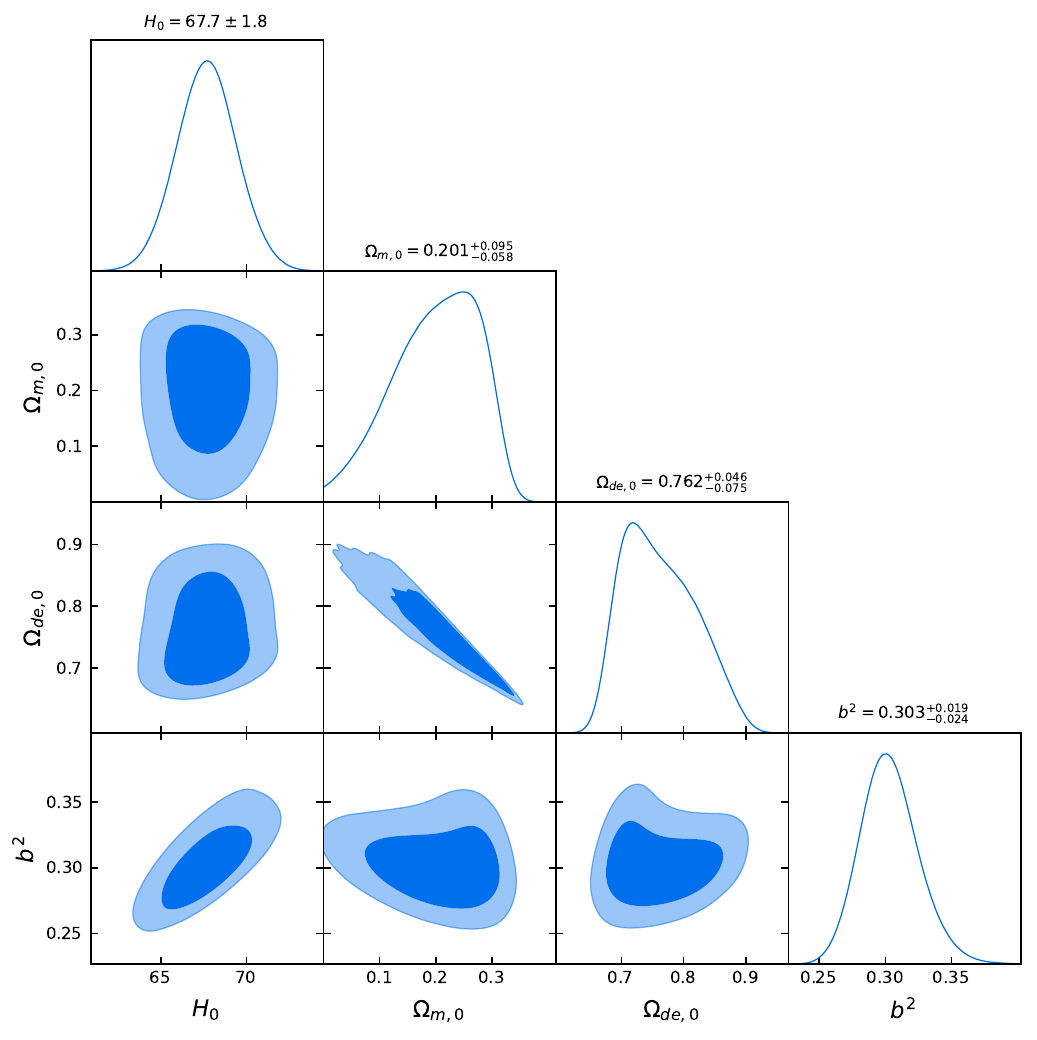}
\caption{\label{Fig0} Confidence contours for the best-fit parameters of the ADEADS model using SN and \mbox{OHD datasets}.}
\end{center}
\end{figure*}

To compare the ADEADS model with the standard $\Lambda$CDM model, we also perform a corresponding MCMC analysis of the standard $\Lambda$CDM model using the same observational datasets. Given the different numbers of parameters in the ADEADS and $\Lambda$CDM models, we employ the Akaike Information Criterion (AIC)~\cite{Akaike1974} and Bayesian Information Criterion (BIC)~\cite{Schwarz1978, Trotta2008} to perform a statistical comparison between these two models, where
\begin{equation}
AIC=\chi^{2}_{min}+2n,
\end{equation}
and
\begin{equation}
BIC=\chi^{2}_{min}+n\ln(N),
\end{equation}
where $n$ and $N$ are the numbers of free parameters in the model and data points, respectively. The constraints on the model parameters for both ADEADS and $\Lambda$CDM are summarized in Table~\ref{Tab0}, which lists their mean values and 1$\sigma$ confidence levels (CLs). Additionally, we present the posterior distributions for the ADEADS model parameters in Figure~\ref{Fig0}.

\begin{table*}
\centering
\fontsize{9}{11}\selectfont
\caption{\label{Tab0} Observational constraints for ADEADS and $\Lambda$CDM models.}
\small
\begin{tabular}{|c|c|c|c|c|c|c|c|}
  \hline
  \hline
  \boldmath{$Model$} & \boldmath{$H_{0}$} &\boldmath{ $\Omega_{m,0}$} & \boldmath{$\Omega_{de,0}$} & \boldmath{$b^{2}$} & \boldmath{$\chi^{2}_{min}$} & \boldmath{$AIC$} & \boldmath{$BIC$}\\
  \hline
  $ADEADS$ & $67.7 \pm 1.8$ & $0.201^{+0.095}_{-0.058}$ & $0.762^{+0.046}_{-0.075}$ & $0.303^{+0.019}_{-0.024}$ & $1418$ & $1430$ & $1462$\\
  \hline
  $\Lambda CDM$ & $67.3 \pm 1.7$ & $0.280^{+0.051}_{-0.026}$ & $0.691^{+0.021}_{-0.027}$ & $-$ & $1418$ & $1428$ & $1455$\\
  \hline
  \hline
  \end{tabular}
\end{table*}

According to the results in Table~\ref{Tab0} and Figure~\ref{Fig0}, the SN and OHD datasets constrain the model parameters of the ADEADS model. The best-fit values are $H_{0}=68.09$, $\Omega_{m,0}=0.285$, $\Omega_{de,0}=0.710$, and $b^{2}=0.292$. This result, corresponding to the constrained values of $H_{0}=67.7 \pm 1.8$, $\Omega_{m,0}=0.201^{+0.095}_{-0.058}$, $\Omega_{de,0}=0.762^{+0.046}_{-0.075}$, and $b^{2}=0.303^{+0.019}_{-0.024}$, differs from the results when using the Hubble radius or the particle horizon as the IR cutoff~\cite{Nakarachinda2022}. Based on these results, both the ADEADS and $\Lambda$CDM models provide an equally good fit, as indicated by their identical $\chi^{2}_{min}$ value of $1418$. Compared to the $\Lambda$CDM model, the AIC of the ADEADS model only increased by $2$. This small difference suggests that the AIC can hardly distinguish which of the two models is better~\cite{Burnham2004}. However, the BIC of the ADEADS model is larger than that of the $\Lambda$CDM model by $7$, indicating strong evidence against the ADEADS model~\cite{Jeffreys1998}. This discrepancy arises because the BIC imposes a much heavier penalty on additional model parameters than the AIC, especially for large datasets.\vspace{-3pt}

\section{Evolution of Universe} \label{sec:4}

To study the universe's evolution in the ADEADS model, {we adopt two sets of initial conditions for comparison: (1) the Planck 2018 results~\cite{Planck2020} with $\Omega_{m,0}=0.3111$, $\Omega_{de,0}=0.6888$, and $H_{0}=67.66 \mathrm{km s}^{-1} \mathrm{Mpc}^{-1}$~\cite{Planck2020}, and (2) the best-fit values constrained by the SN Ia and OHD datasets in this work, with $\Omega_{m,0}=0.285$, $\Omega_{de,0}=0.710$, and $H_{0}=68.09 \mathrm{km s}^{-1} \mathrm{Mpc}^{-1}$. After solving Equations~(\ref{Omm})--(\ref{OHT}) numerically for different values of $b^{2}$ and $\alpha$, we obtain the evolutionary curves of $\Omega_{de}$, $\xi$, $\omega_{de}$, and $q$, as shown in Figures~\ref{Fig1}--\ref{Fig3}. 

The left panel of Figure~\ref{Fig1} illustrates the evolution of $\Omega_{de}$ for different parameter combinations $(b^{2}, \alpha)$. As the redshift decreases, $\Omega_{de}$ asymptotically approaches $1$, regardless of the initial parameter choices. This behavior indicates that the late-time dominance of dark energy is an attractor solution in the ADEADS model, implying that ADEADS dominates the late-time evolution of the universe. The right panel of Figure~\ref{Fig1} shows the evolution of $\Omega_{r}$, $\Omega_{m}$, and $\Omega_{de}$; the solid line denotes the case for the best-fit values, while the dashed line represents the case for $b^{2}=0.2, \alpha=2.5$. The curves successfully reproduce the standard succession of cosmic eras, from early radiation domination to matter domination, and finally to the dark energy-dominated epoch. This evolution confirms the model's consistency with the entire cosmic history.

Figure~\ref{Fig2} illustrates the evolution of $\xi$ as a function of redshift $z$, demonstrating that $\xi>0$ across the entire cosmic history. Both panels show that $\xi$ emerges from $0$ in the early universe, undergoes a characteristic peak, and subsequently decays back to $0$ in the future. The left panel, with a logarithmic scale in $\ln(1+z)$, effectively captures the long-term behavior across all epochs, while the right panel's linear scale in $1+z$ offers a clear view of the evolution of $\xi$ in the more recent universe.

The left panel of Figure~\ref{Fig3} indicates that ADEADS behaves as quintessence: $\omega_{de}$ starts at $-1$, evolves away from it, and eventually approaches $-1$ again, mimicking a cosmological constant at late times. The right panel of Figure~\ref{Fig3} demonstrates that late-time acceleration is realized, with a viable transition redshift ($0.48\leq z_{t} <1$), for both of our best-fit values and cases using the Planck 2018 result as initial conditions with $b^{2}=0.2,0.4,0.6,0.8$. In addition, for the case of the best-fit values, the deceleration parameter $q$ deviates significantly from the other cases at $z>6$, i.e., in the early universe.

\begin{figure*}[htp]
\begin{center}
\includegraphics[width=0.452\textwidth]{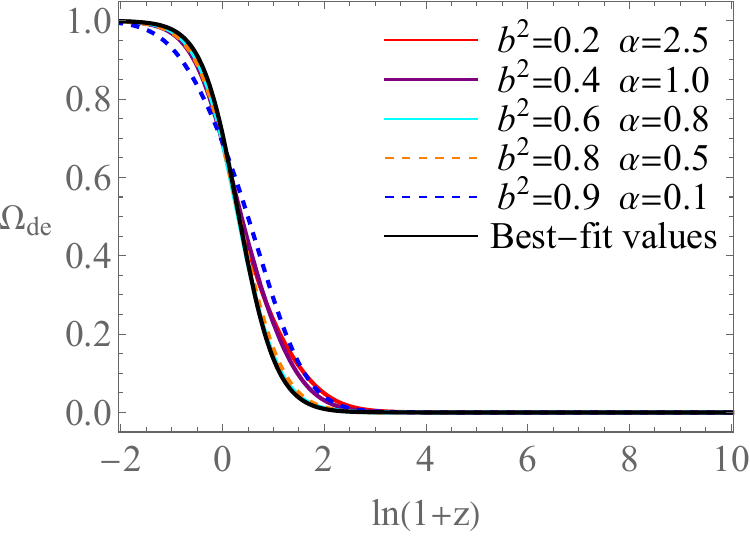}
\includegraphics[width=0.435\textwidth]{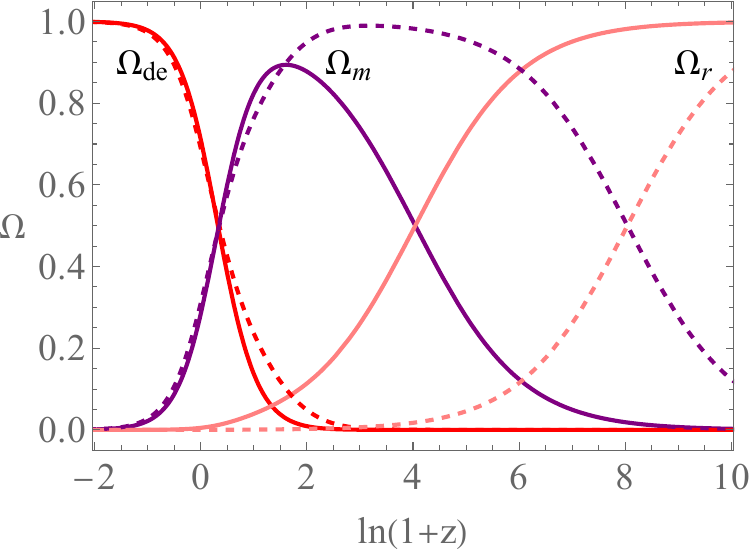}
\caption{\label{Fig1} Evolution curves of $\Omega_{de}$ and $\Omega$ versus redshift parameter $\ln(1+z)$.}
\end{center}
\end{figure*}

\begin{figure*}[htp]
\begin{center}
\includegraphics[width=0.452\textwidth]{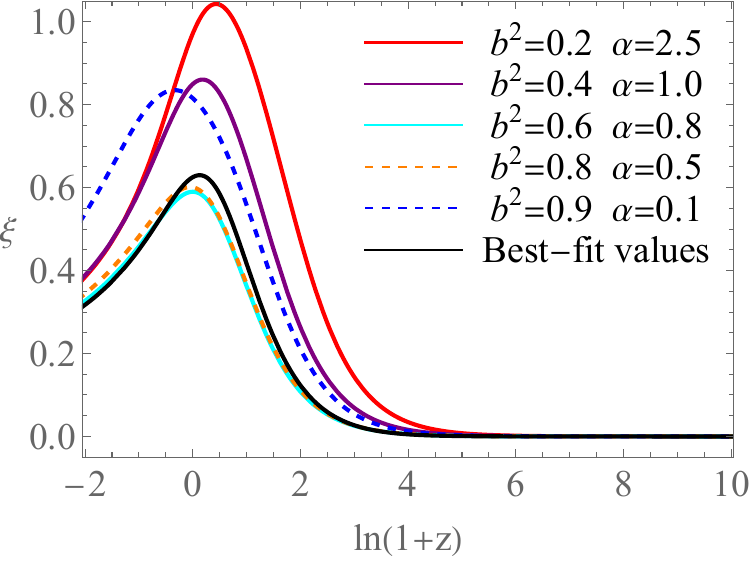}
\includegraphics[width=0.452\textwidth]{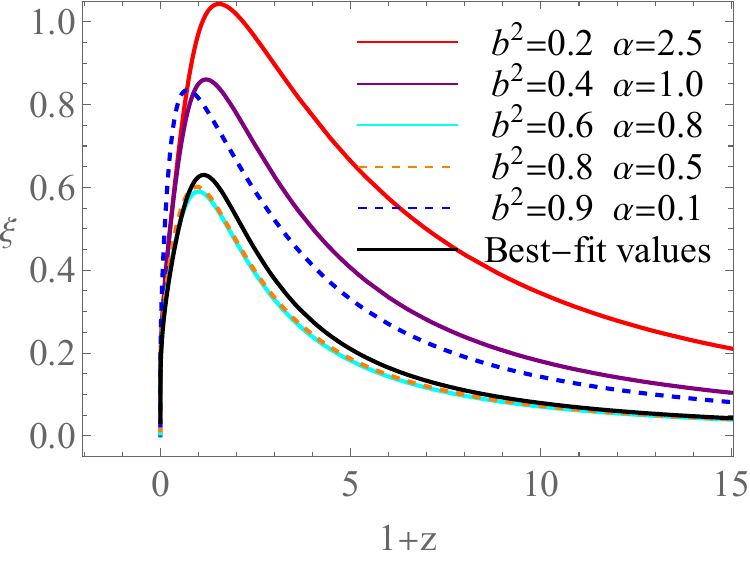}
\caption{\label{Fig2} Evolution curves of $\xi$ versus redshift parameter $\ln(1+z)$ and $1+z$.}
\end{center}
\end{figure*}

\begin{figure*}[htp]
\begin{center}
\includegraphics[width=0.452\textwidth]{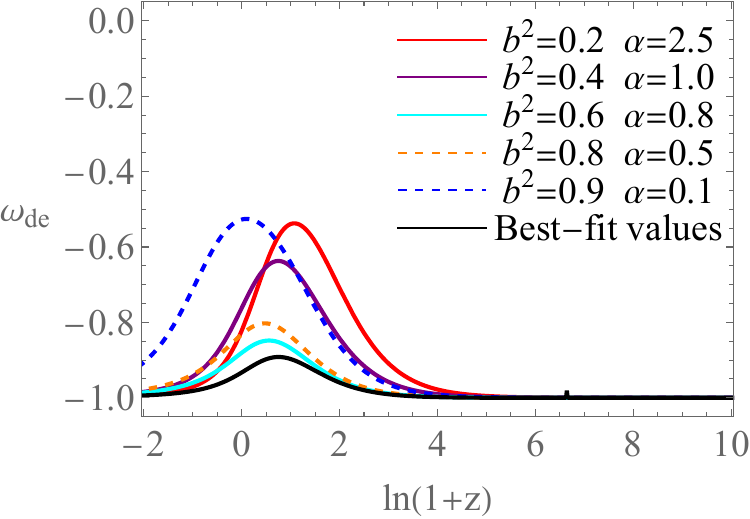}
\includegraphics[width=0.435\textwidth]{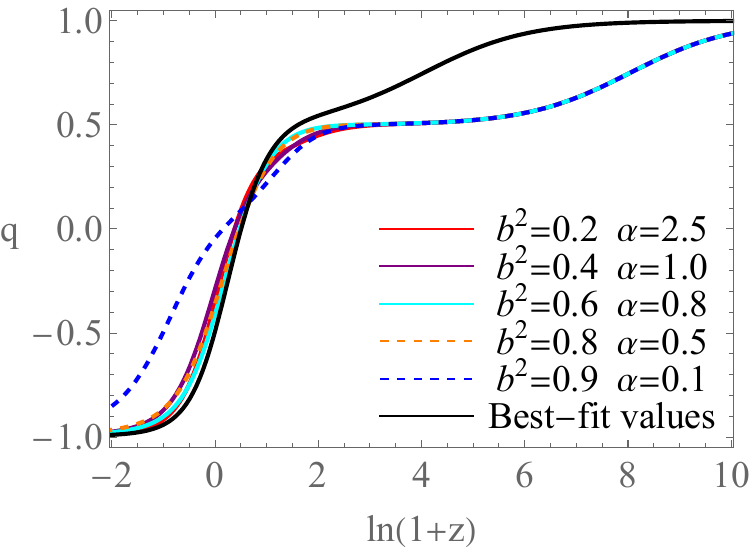}
\caption{\label{Fig3} Evolution curves of $\omega_{de}$ and $q$ versus redshift parameter $\ln(1+z)$.}
\end{center}
\end{figure*}

Overall, as both $q$ and $\omega_{de}$ asymptotically approach $-1$, the ADEADS model naturally explains the present cosmic acceleration, reproduces a $\Lambda$CDM-like late-time phase, and provides a consistent description of the universe's full evolutionary history, as is shown in the right panel of Figure~\ref{Fig1}.

\section{Hubble Diagram} \label{sec:5}

In the previous section, we analyzed the universe's evolution in the ADEADS model for selected parameter sets. Here, we test their consistency with Hubble observational data. Figure~\ref{Fig3z} presents the evolution of the Hubble parameter $H(z)$, where the error bars correspond to observational measurements~\cite{Akhlaghi2018, Cao2021}. The black solid line represents the evolutionary curves produced by the best-fit values, which overlaps with the black dashed line representing the standard $\Lambda$CDM model. For the cases that adopt the Planck 2018 results as initial conditions, the predicted curves deviate slightly from the $\Lambda$CDM model while remaining in good agreement with the data. Furthermore, consistency with observations requires that a larger value of $b^{2}$ be compensated by a smaller value of $\alpha$.
\begin{figure*}[htp]
\begin{center}
\includegraphics[width=0.5\textwidth]{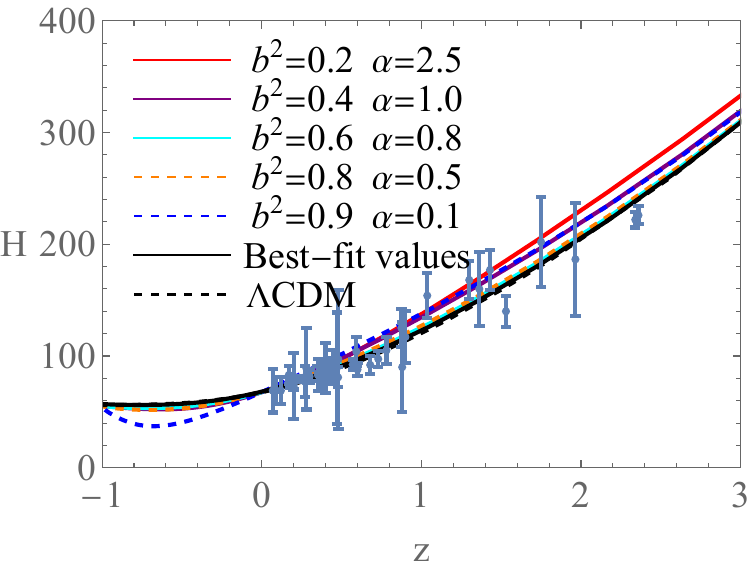}
\caption{\label{Fig3z} Evolution curves of $H$.}
\end{center}
\end{figure*}

A turning point in $H(z)$ was previously found in the HDE model with the future event horizon as the IR cutoff~\cite{Colgain2021}, but was not found in the Barrow holographic dark energy model~\cite{Huang2021}. In the ADEADS model, such a turning point appears in the specific case of $b^{2}=0.9,\alpha=0.1$, whereas it is absent for the best-fit values and all other parameter sets analyzed in this work.

\section{Dynamical Analysis} \label{sec:6}

In the section preceding, we have demonstrated that the ADEADS model can reproduce the entire cosmic evolution and mimic a cosmological constant. Here, we further analyze its dynamical behavior using the dynamical system approach~\cite{Bahamonde2018, Wu2010, Huang2019a, Huang2021, Wu2007, Wu2008, Dutta2016, Dutta2017, Dutta2019, Huang2025, Huang2021a, HuangY2015, Chen2009, Chatterjee2025}. Critical points of the autonomous system are obtained by solving
\begin{equation}
\Omega_{m}'=\Omega_{de}'=\xi'=0.
\end{equation}
For systems~(\ref{Omm})--(\ref{OHT}), seven solutions arise, but one is excluded due to the physical requirement $\xi \geq 0$. The remaining six critical points are listed in Table~\ref{Tab1}. Among them, points $P_{1}$, $P_{2}$, and $P_{3}$ require $\xi=0$, a condition that holds asymptotically through most of the universe's evolution, as shown in Figure~\ref{Fig2}. $P_{1}$ corresponds to the radiation-dominated decelerating epoch, $P_{2}$ to the matter-dominated decelerating epoch, and $P_{3}$ to an accelerating epoch dominated by ADEADS, which behaves like a cosmological constant since $\omega_{de}=-1$. Points $P_{4}$, $P_{5}$, and $P_{6}$ depend on the parameter $b^{2}$: for $b^{2}=\frac{4}{9}$, $P_{4}$ describes an ADEADS-dominated decelerating epoch, while for small $b^{2}$, it reduces to matter domination since $\omega_{de}=0$; for $b^{2}=\frac{1}{4}$, $P_{5}$ corresponds to ADEADS domination but approaches radiation behavior since $\omega_{de}=\frac{1}{3}$ under small $b^{2}$.

\begin{table*}
\centering
\fontsize{9}{11}\selectfont
\caption{\label{Tab1} Critical points and stability conditions of ADEADS.}
 \begin{tabular}{|c|c|c|c|c|c|c|c|}
  \hline
  \hline
  \boldmath{$Label$} & \boldmath{$Points (\Omega_{m}, \Omega_{de}, \xi)$} & \boldmath{$\Omega_{r}$} & \boldmath{$\omega_{de}$} & \boldmath{$q$} & \boldmath{$Eigenvalues$} & \boldmath{ $Conditions$} & \boldmath{$Points$} \\
  \hline
  $P_{1}$ & $(0,0,0)$ & $1$ & $-1$ & $1$ & $(4,2,1)$ & $Always$ & $Unstable \ point$\\
  \hline
  $P_{2}$ & $(1,0,0)$ & $0$ & $-1$ & $\frac{1}{2}$ & $(3,\frac{3}{2},-1)$ & $Always$ & $Saddle \ point$\\
  \hline
  $P_{3}$ & $(0,1,0)$ & $0$ & $-1$ & $-1$ & $(-4,-3,0)$ & $Always$ & $Stable \ point$ \\
  \hline
  $P_{4}$ & $(1-\frac{9}{4}b^{2},\frac{9}{4}b^{2},\frac{3}{2})$ & $0$ & $0$ & $\frac{1}{2}$ & $(3,-1,-\frac{3}{2}+\frac{27}{8}b^{2})$ & $Always$ & $Saddle \ point$\\
  \hline 
  $P_{5}$ & $(0,4 b^{2},2)$ & $1-4 b^{2}$ & $\frac{1}{3}$ & $1$ & $(4,1,-2+8 b^{2})$ & $b^{2}<\frac{1}{4}$ & $Saddle \ point$\\
  \hline
  $P_{6}$ & $(0,1,\frac{1}{\sqrt{b^{2}}})$ & $0$ & $-1+\frac{2}{3\sqrt{b^{2}}}$ & $-1+\frac{1}{\sqrt{b^{2}}}$ & $(\frac{2}{\sqrt{b^{2}}},-4+\frac{2}{\sqrt{b^{2}}},-3+\frac{2}{\sqrt{b^{2}}})$ & $\frac{4}{9}<b^{2}<1$ & $Saddle \ point$\\
  \hline
  \hline
  \end{tabular}
\end{table*}

Linearizing Equations~(\ref{Omm})--(\ref{OHT}) yields the stability properties summarized in Table~\ref{Tab1}. Points with all negative eigenvalues are stable attractors, all positive are unstable, and mixed signs are saddles. $P_{3}$ is non-hyperbolic due to a vanishing eigenvalue. To analyze its stability, numerical methods can be employed to investigate the asymptotic behavior near this critical point~\cite{Dutta2016, Dutta2017, Dutta2019}. In Figure~\ref{Fig4}, we have depicted the time evolution of trajectories projected onto the $\Omega_{m}$, $\Omega_{de}$, and $\xi$-axis for critical point $P_{3}$. From this figure, under initial perturbations, the trajectories converge toward point $P_{3}$ without divergence, thereby confirming that it is a stable attractor.

\begin{figure*}[htp]
\begin{center}
\includegraphics[width=0.33\textwidth]{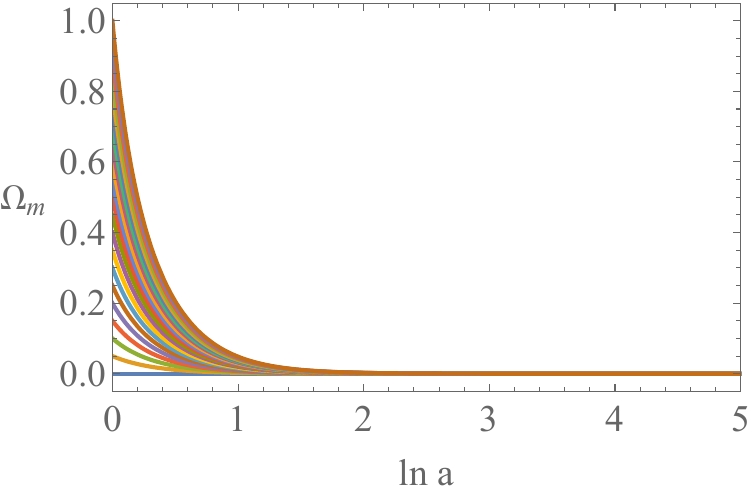}
\includegraphics[width=0.33\textwidth]{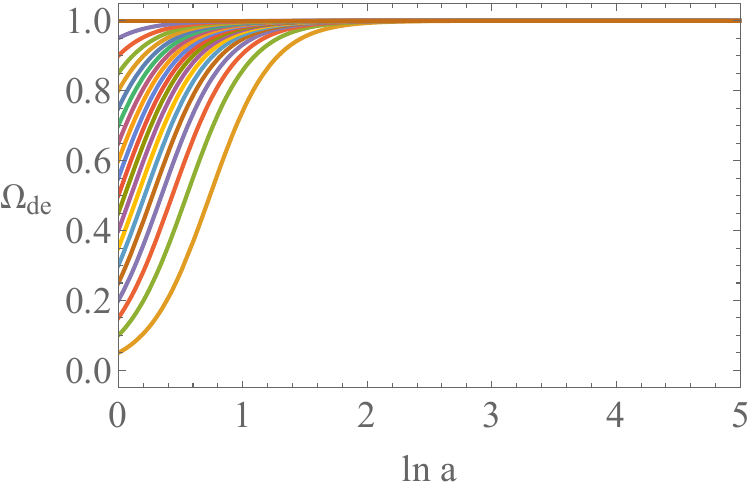}
\includegraphics[width=0.32\textwidth]{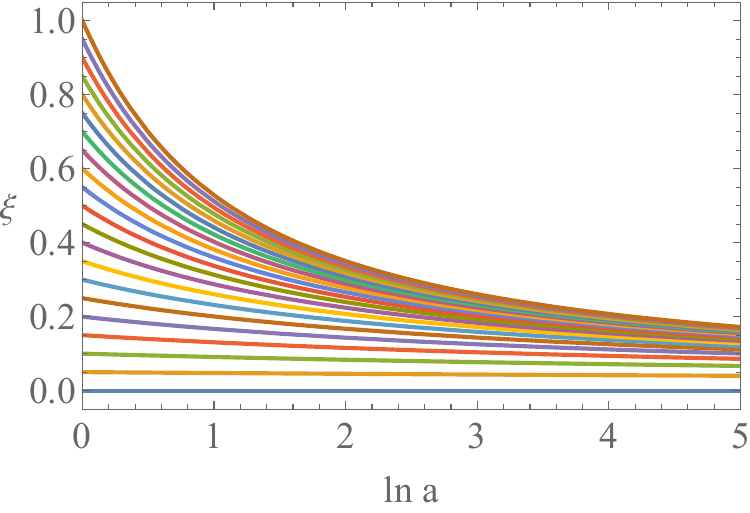}
\caption{\label{Fig4} Time evolution of trajectories projected on $\Omega_{m}$, $\Omega_{de}$, and $\xi$-axis for $P_{3}$. These panels are plotted for $b^{2}=0.2$.}
\end{center}
\end{figure*}

Table~\ref{Tab1} further indicates that $P_{1}$ is unstable, $P_{3}$ is stable, and the others act as saddle points. We have plotted the phase diagram with critical points and evolutionary trajectories in Figure~\ref{Fig5}; the left panel displays the $(\Omega_{m}$, $\Omega_{de}$, $\xi$) space configuration, while the right panel shows the projection onto the ($\Omega_{m}$, $\Omega_{de}$) plane. The stability analysis of critical points presented in Table~\ref{Tab1} and Figure~\ref{Fig5} indicates that the universe will ultimately enter a late-time acceleration epoch dominated by ADEADS, effectively emulating the cosmological constant. If the universe initially evolves from the radiation-dominated epoch $(P_{1})$, it can shift to the pressureless matter-dominated epoch $(P_{2})$, eventually entering the late-time acceleration era dominated by ADEADS $(P_{3})$; this scenario comprehensively describes the whole evolution trajectory of the universe and is graphically presented by the red curves \mbox{in Figure~\ref{Fig5}.}

From Figure~\ref{Fig5}, we can see that three important critical points ($P_{1},P_{2},P_{3}$) exist, which describe the entire evolution of the universe and are located on the $\xi=0$ plane. To intuitively investigate the dynamics of these points, the phase diagram of ($\Omega_{m}$, $\Omega_{de}$) on the $\xi=0$ plane is shown in Figure~\ref{Fig6}, which is plotted for $b^{2}=0.2$; the variation in $b^{2}$ will have no impact on the dynamic behavior of this system. In this figure, the red curve uses the Planck 2018 results as the initial conditions, and the black curve uses our best-fit values. The magenta dots mark the current values from the best-fit values, while the purple dots mark those from the Planck 2018 results. Thus, this model is suitable for describing the universe's evolution, and the universe will ultimately enter into a phase described by the cosmological constant $\Lambda$ since ADEADS can exhibit cosmological constant behavior during late-time cosmic expansion.

\begin{figure*}[htp]
\begin{center}
\includegraphics[width=0.44\textwidth]{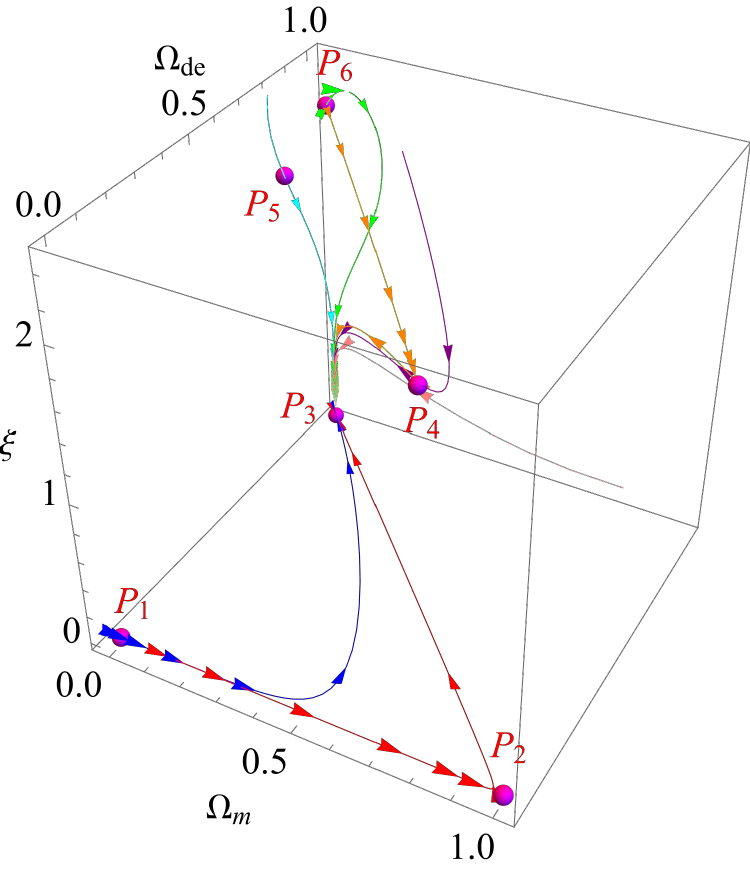}
\includegraphics[width=0.45\textwidth]{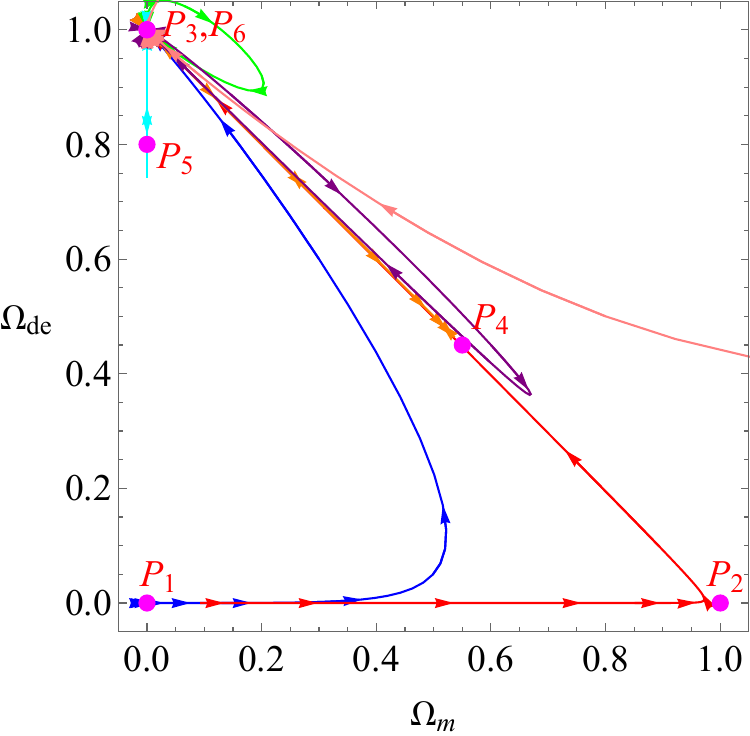}
\caption{\label{Fig5} Phase diagram with critical points and evolutionary trajectories. The left panel is plotted for the $(\Omega_{m}$, $\Omega_{de}$, $\xi$) space, while the right one is for the ($\Omega_{m}$, $\Omega_{de}$) plane.}
\end{center}
\end{figure*}

\begin{figure*}[htp]
\begin{center}
\includegraphics[width=0.5\textwidth]{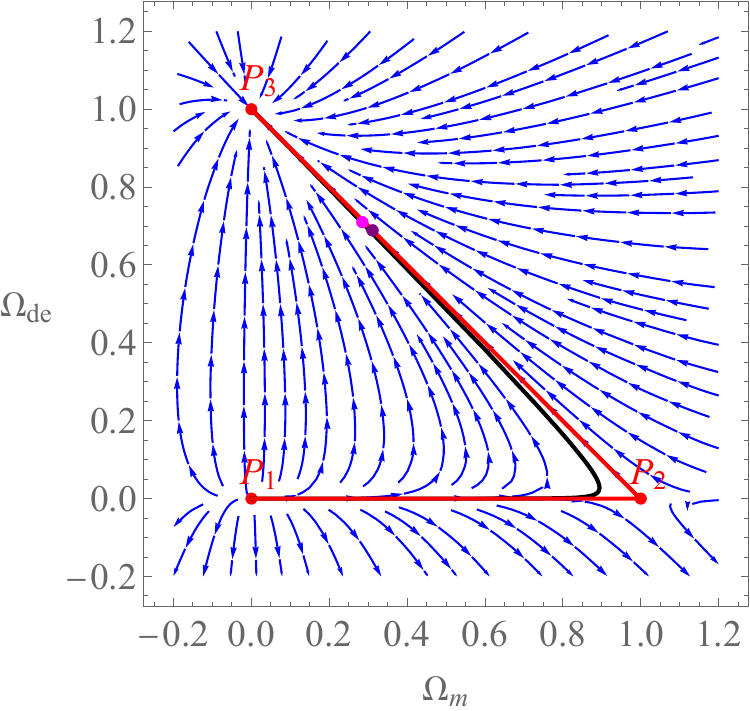}
\caption{\label{Fig6} Phase diagram of ($\Omega_{m}$, $\Omega_{de}$) on the $\xi=0$ plane.}
\end{center}
\end{figure*}

\section{Statefinder Analysis} \label{sec:7}

Based on the results of the previous section, we have shown that the ADEADS model can reproduce the entire cosmic evolution and has an attractor that behaves like the cosmological constant. To further differentiate ADEADS from the standard $\Lambda$CDM scenario, we adopt the statefinder diagnostic pairs $\{r, s\}$ and $\{r, q\}$ that were originally introduced by Sahni {et~al.}~\cite{Sahni2003}. 

The statefinder parameters $r$ and $s$, defined as geometric diagnostics derived exclusively from the cosmic scale factor $a$, are expressed as~\cite{Sahni2003, Wu2005}
\begin{equation}
r=\frac{\dddot{a}}{a H^{3}}, \qquad s=\frac{r-1}{3(q-\frac{1}{2})}.
\end{equation}
Through differentiation of Equation~(\ref{HH2}), the statefinder parameters $r$ and $s$ can be calculated according to $\Omega_{m}'$, $\Omega_{de}'$, and $\xi'$. Thus, by solving Equations~(\ref{Omm})--(\ref{OHT}) numerically, we can obtain the evolutionary curves of the universe in the $\{r, s\}$ and $\{r, q\}$ planes. 

In the left panel of Figure~\ref{Fig7}, we depict some examples of the evolutionary curves for the statefinder diagnostic pairs $\{r, s\}$. The green line corresponds to the $\Lambda$CDM model, with its fixed point $(0,1)$ marked by a green dot, while other dots represent current values from respective trajectories. The evolution curves for these examples deviate from the $\Lambda$CDM model in the late-time evolution, enter the quintessence region, and eventually evolve to the $\Lambda$CDM fixed point $(0,1)$. However, the evolutionary curves for the case for the best-fit values shows a significant deviation from the $\Lambda$CDM model. In the right panel of Figure~\ref{Fig7}, we plot another statefinder diagnostic pair in the $\{r, q\}$ plane, where the de Sitter expansion fixed point $(-1,1)$ shown by the green dot, and the standard cold dark matter fixed point $(0.5,1)$, corresponds to the magenta dot. The evolutionary curves based on the Planck 2018 results as initial conditions originate from the standard cold dark matter fixed point $(0.5,1)$ and eventually converge toward the de Sitter expansion fixed point $(-1,1)$. In contrast, the curves for the best-fit values not only show a significant deviation from the $\Lambda$CDM model but also fail to originate from the standard cold dark matter fixed point. Thus, the ADEADS model can be differentiated from the standard $\Lambda$CDM model by the statefinder diagnostic pairs $\{r, s\}$ and $\{r, q\}$.

\begin{figure*}[htp]
\begin{center}
\includegraphics[width=0.45\textwidth]{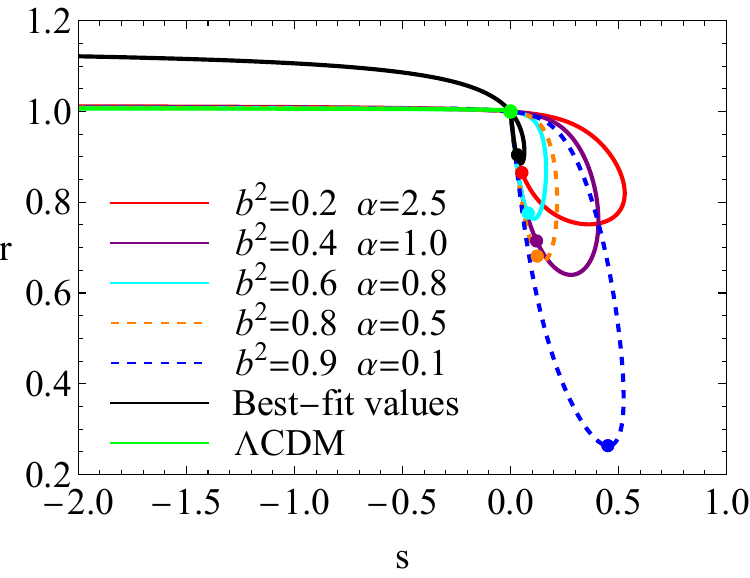}
\includegraphics[width=0.45\textwidth]{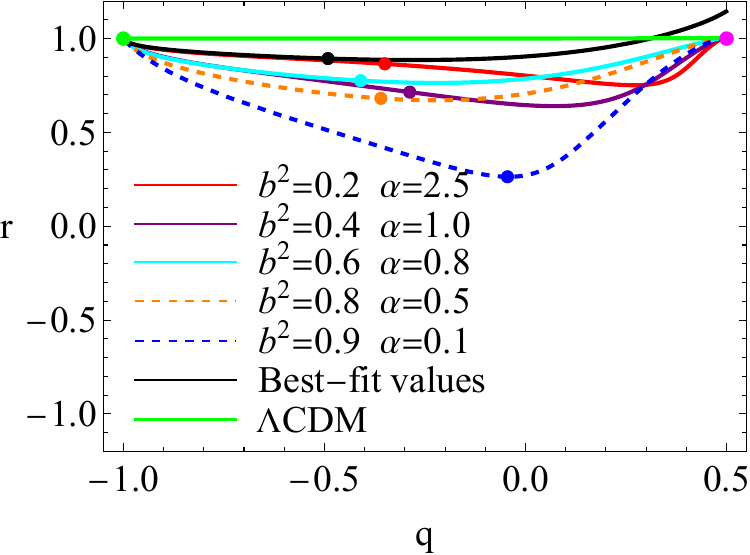}
\caption{\label{Fig7} Statefinder diagnostics in the $\{r, s\}$ and $\{r, q\}$ planes. The dots represent the corresponding current values.}
\end{center}
\end{figure*}

\section{Conclusions} \label{sec:8}

Using the entropy derived from AdS black holes and adopting the Hubble horizon as the infrared cutoff, a new HDE model has been constructed, which can achieve the late-time accelerated expansion of the universe. In this paper, by choosing the age of the universe as the IR cutoff, we construct the ADEADS model and constrain its parameters using SN Ia and OHD datasets. We find that the Akaike Information Criterion cannot effectively distinguish this model from the standard $\Lambda$CDM model, and the present value of Hubble constant $H_{0}$ and the model parameter $b^{2}$ are constrained to $H_{0}=67.7 \pm 1.8$ and $b^{2}=0.303^{+0.019}_{-0.024}$, respectively. We then analyze the cosmic evolution within the ADEADS model using two sets of initial conditions: the Planck 2018 results and the best-fit values obtained in this work. The results show that the ADEADS model, which can depict the entire evolutionary history of the universe dominated by radiation, pressureless matter, and ADEADS itself, achieves late-time acceleration by mimicking the cosmological constant $\Lambda$, as evidenced by $\omega_{de}$ and $q$ approaching $-1$. When these cases are examined against the Hubble diagram, we find that the turning point in the Hubble parameter is absent in the ADEADS model for specific parameter choices. Then, applying dynamical analysis techniques to the ADEADS model, we find that the universe will ultimately enter into an epoch characterized by the standard $\Lambda$CDM model since the attractor in the ADEADS model represents an epoch described by the cosmological constant $\Lambda$. In order to distinguish the ADEADS model from the standard $\Lambda$CDM model, we adopt the statefinder analysis method and find that the evolution curves for these examples deviate from the $\Lambda$CDM model in the late-time evolution, which indicates that the ADEADS model can be distinguished from the standard $\Lambda$CDM model. Thus, with the age of the universe as the IR cutoff, the ADEADS model successfully describes the entire cosmic evolutionary history, including the late-time acceleration. It asymptotically approaches the standard $\Lambda$CDM model while remaining observationally distinguishable from it in the late universe.

\acknowledgments{This work was supported by the National Natural Science Foundation of China under Grant Nos. 12405081, 12265019, and 11865018. }

\end{document}